\documentstyle[twoside,fleqn,espcrc2]{article}
\newcommand{\AmS}{{\protect\the\textfont2
  A\kern-.1667em\lower.5ex\hbox{M}\kern-.125emS}}

\title{The Large Mixing of the Pseudoscalars in QCD and Large Flavor Mixing
of Neutrinos}
                                
\author{H. Fritzsch\address{Sektion Physik der Universit\"at M\"unchen,\\
        Theresienstrasse 37, D--80333 M\"unchen}
        }
\begin{document}
\begin{abstract}
In analogy to the mixing pattern of the pseudoscalar mesons in QCD we discuss
the mixing of massive neutrinos. Unlike the quarks flavor mixing angles the
leptonic mixing angles are large, nearly maximal. The three massive neutrinos
are nearly degenerate.
\end{abstract}

\maketitle
\setcounter{page}{1}
Unlike the quarks, gluons are not very conspicuous in strong
interaction physics. The quarks exhibit their flavor quite openly, but
gluons act in the background and are difficult to identify.
There are plenty of ``smoking guns'' for the quark degrees of freedom, but it
is hard to identify one for the glue, with one exception: The mass and mixing
pattern of the $0^{-+}$--mesons. In fact, the peculiarities of these mesons,
especially the $SU(3)$--singlet mesons $\eta '$, were the primary reason to
construct the theory of quarks and gluons more than 25 years
ago \cite{frige}.\\

In QCD we are able to turn off the masses of the $u, d$ and $s$--quarks,
i. e. consider the case of an exact $SU(3)_L \times SU(3)_R$--symmetry. In
this limit we are dealing with eight massless Goldstone bosons (pions,
kaons and $\eta $) and a massive $\eta '$--meson. Due to the QCD anomaly the
latter has a mass of the order of 1 GeV. The mass--squared matrix for the
neutral bosons $\left( \pi^{\circ}, \eta, \eta' \right)$ can be written as
\begin{equation}
M^2 = m^2 \left( \eta ' \right) \left( \begin{array}{ccc}
                                        0 & 0 & 0\\
                                        0 & 0 & 0\\
                                        0 & 0 & 1
                                        \end{array} \right) \, .
\end{equation}
By a simple orthogonal transformation this rank--one matrix can be
transformed into a matrix $M^2_d$, in which all entries are equal:
\begin{eqnarray}
U^{-1} \, M^2 \, U & = & M^2_d = \frac{1}{3} m^2 \left( \eta ' \right)
\left( \begin{array}{ccc}
       1 & 1 & 1 \\
       1 & 1 & 1 \\
       1 & 1 & 1 
       \end{array} \right)    \nonumber \\
U & = & \left( \begin{array}{ccc} 
               \frac{1}{\sqrt{2}} & - \frac{1}{\sqrt{2}} & 0\\
               \frac{1}{\sqrt{6}} &   \frac{1}{\sqrt{6}} & \frac{-2}{\sqrt{6}}\\
               \frac{1}{\sqrt{3}} &   \frac{1}{\sqrt{3}} & \frac{1}{\sqrt{3}}
               \end{array} \right)
\end{eqnarray}
It is interesting to consider the basis, in which the mass--squared matrix
has this ``democratic'' form, in more detail. The meson wave functions are
given by:
\begin{eqnarray}
\pi _0 & = & \frac{1}{\sqrt{2}} \, \left( \bar u u - \bar d d \right)
\nonumber \\
\eta & = & \frac{1}{\sqrt{6}} \, \left( \bar u u + \bar d d - 2 \bar s s \right)
\nonumber \\
\eta' & = & \frac{1}{\sqrt{3}} \left( \bar u u + \bar d d + \bar s s \right)
\end{eqnarray}
The unitary transformation given above is just the one which describes the
neutral pseudoscalars in terms of ``pure'' $\left( \bar q q \right)$--states.
Thus the ``democratic'' mass--squared matrix arises in the basis of the
$\left( \bar q q \right)$-- states. In QCD this is easily understood as the
consequence of the gluonic interaction. The latter allows
transitions between $\bar u u$ and $\bar d d$, $\bar u u$ and $\bar s s$ etc.,
universal in strength, which would be absent in a free quark model. The
universality of those transitions is described by the democratic form of the
mass matrix in the flavor basis. Their coherent action leads to the mass of
the $\eta '$--meson, while the same coherence forces the
$\pi^{\circ }$ and $\eta$--mesons to stay massless. Their masses are due to
the violation of the chiral symmetry which in particular generates small
violations of the democratic structure of the mass matrix.

Qualitatively the masses of the quarks and of the charged leptons show a
similar pattern, when viewed in each flavor channel: the members of the first
fmaily $(u, d, e^-)$ are nearly massless, the members of the second family
$(c, s, \mu ^-)$ have masses which are relatively small compared to the masses
of the members of the third family $(t, b, \tau ^-)$. These features are
valid for the charged fermions. There is no reason that the neutrinos show a
similar pattern. Both in case of the charged leptons and of the quarks the
contribution of the first family to the total mass in the corresponding
flavor channel is almost negligible.

For the discussion of flavor mixing it is often useful to treat the fermion
masses as parameters, which can be manipulated in a similar way as the
masses of the pseudoscalar masses in QCD and in particular set to zero or
infinity. Obviously the physics of the leptons and quarks will not be changed
significantly, if we set the masses of the members of the first and second
family to zero. The departure from the real world will only be about 5\%.
Or course, due to our ignorance about the origin of the fermion masses we do
not know whether such a manipulation of these masses is indeed possible.
Within the framework of the standard model it is, of course, easy to make
such changes just be modifying the coupling parameters describing the
interaction of the fermions with the scalar field.

If the masses of the first two families vanish, the mass matrices of the
fermions become matrices of rank one:
\begin{equation}
M = C \left( \begin{array}{ccc}
             0 & 0 & 0 \\
             0 & 0 & 0\\
             0 & 0 & 1 \end{array} \right) \, .
\end{equation}
In the limit there appears to be a mass gap $C$, given by the mass of the
$t$--quark, the $b$--quark or the $\tau $--lepton respectively, just like
the mass gap in the chiral limit of QCD, given by the $\eta '$--mass.
By a suitable orthogonal transformation the mass matrix $M$ can be brought
into a form $\bar M$, in which all matrix elements are identical:
\begin{equation}
\bar M = \frac{C}{3} \left( \begin{array}{ccc}
                             1 & 1 & 1 \\
                             1 & 1 & 1 \\
                             1 & 1 & 1 \end{array} \right) \, .
\end{equation}
Such a mass matrix, which is often called a democratic mass matrix, also has
rank one. As far as the charged leptons and the quarks are concerned, we can
speak either of a hierarchy basis $(M)$ or of a democratic basis $(\bar M)$.
Both are, of course, equivalent. However, in the democratic basis one
realizes a new symmetry described by the discrete group
$S(3)_L \times S(3)_R$. Any permutation of two rows or columns of the mass
matrix $\bar M $ leaves
$\bar M$ invariant.

In analogy to the pseudoscalar mesons let me write down
the mass eigenstates of charged leptons in the democratic limit in terms of
the eigenstates $l_1, l_2$ and $l_3$ of the democratic
symmetry \cite{frihp}.
\begin{eqnarray}
e & = & \frac{1}{\sqrt{2}} \left( l_1 -l_2 \right) \, , \nonumber \\
\mu & = & \frac{1}{\sqrt{6}} \left( l_1 + l_2 - 2l_3 \right) \, , \nonumber \\
\tau & = & \frac{1}{\sqrt{3}} \left( l_1 + l_2 + l_3 \right) \, .
\end{eqnarray}

Similar relations can be written down for the quarks. In reality the
democratic symmetry is not exact, but broken by small terms. These symmetry
breaking effects have been discussed some time ago by a number of authors,
and we shall refer to the literature \cite{frihp}, \cite{harari}.

It is interesting to discuss the description of the flavor mixing in the
context of the democratic symmetry and its violation. In the limit of the
democratic symmetry one expects for the quarks that no flavor mixing is
present. In other words, the flavor mixing angles must be related to the
violation of the symmetry and in particular to the masses of the first two
families, or rather to the mass ratios of the masses of the first two
families and the mass of the corresponding representative of the third
family. An interesting way of describing the flavor mixing would be one in
which
the parameters for the flavor mixing, e.\ g.\ the flavor mixing angles, are
smooth functions of the symmetry breaking parameters \cite{frixi}.

Since the mass spectrum of the charged leptons exhibits a similar
hierarchical pattern as the quarks, it is most natural to suppose that the
matrix structure and the texture properties of the charged lepton mass
matrix is analogous to those of the quark mass matrices.
The question arises whether the neutrinos also exhibit a hierarchical mass
pattern. It may well be that the neutrino masses are not hierarchical at all.
If they were, we could write the neutrino mass matrix in analogy to the
charged lepton matrix in the democratic basis as follows:
\begin{equation}
M (\nu ) = \frac{C_{\nu }}{3} \left( \begin{array}{ccc}
                                        1 & 1 & 1\\
                                        1 & 1 & 1\\
                                        1 & 1 & 1 \end{array} \right)
+ \Delta M (\nu ) \, ,
\end{equation}
where $\Delta M ( \nu )$ denotes the correction. The constant $C_{\nu } $
describes the strength of the pairing--force term in the neutrino channel.
The magnitude of this term is related to the mass of the heaviest neutrino
and could be at most about 30 eV, according to astrophysical constraints,
i. e. it must be about eight orders of
magnitude smaller than the charged lepton term $(C_l = m_{\tau })$. It is
hard to believe that the ratio $C_{\nu } / C_l$ is simply a tiny number by
accident. It would be much more natural to suppose that the leading
pairing--force term is completely absent in the neutrino channel. This
possibility was discussed in \cite{frixi1}. The absence of the leading
pairing force term for the neutrinos would have drastic consequences for the
mixing pattern of the leptons. An interesting possibility is that the
eigenstates of the democratic symmetry for the neutrinos are identical to
the mass eigenstates. We write down the following mass matrices for the
charged leptons and the neutrinos as follows:
\begin{eqnarray}
M (l^-) & = & C_l \left( \begin{array}{ccc}
                        1 & 1 & 1\\
                        1 & 1 & 1\\
                        1 & 1 & 1 \end{array} \right) +
                \left( \begin{array}{ccc}
                        \delta _l & 0       & 0 \\
                        0         & \rho _l & 0 \\
                        0         & 0       & \varepsilon _l
                        \end{array} \right) \, , \nonumber \\
M (\nu ) & = & 0 + \left( \begin{array}{ccc}
                          \delta _{\nu } &     0        & 0 \\
                           0             & \rho _{\nu } & 0 \\
                           0             &     0        & \varepsilon_{\nu}
                           \end{array} \right) \, .
\end{eqnarray}
Obviously large mixing phenomena are generated due to the absence of the
pairing--force term for the neutrinos. The electroweak doublets of leptons
can be written as:
\begin{equation}
\left( \begin{array}{ccc}
        \nu_1 & \nu_2 & \nu_3 \\
        l_1   & l_2   & l_3 \end{array} \right) \, .
   \end{equation}
Here the upper components, the neutrino states, are the mass eigenstates
while their electroweak partners are identical to the democratic eigenstates
$l_i$. We can also perform a unitary transformation and write down the mass
eigenstates for the charged leptons, neglecting the small breaking terms for
the democratic symmetry, and obtain:
\begin{displaymath}
\left( \begin{array}{c}
        \frac{1}{\sqrt{2}} \nu_1 - \nu_2 \\
         e^- \end{array} \right)
\left(  \begin{array}{c}
        \frac{1}{\sqrt{6}} \nu _1 + \nu_2 - 2 \nu _3 \\ 
        \mu ^- \end{array} \right) 
\end{displaymath}
\begin{equation}
\left( \begin{array}{c}       
       \frac{1}{\sqrt{3}} \left( \nu_1 + \nu_2 + \nu_3 \right)\\
       \tau ^- \end{array} \right) \, .
\end{equation}
We describe the leptonic flavor mixing matrix as follows, in analogy to the
quark sector \cite{frixi2}:
\begin{displaymath}
V_l = \left( \begin{array}{ccc} 
                 c_{\nu } & s_{\nu } & 0\\
                -s_{\nu}  & c_{\nu}  & 0\\
                   0      &   0      & 1 \end{array} \right)
          \left( \begin{array}{ccc}
                 e^{i \psi} & 0 & 0 \\
                 0          & c & s \\
                 0          & -s & c \end{array} \right) \\
\end{displaymath}
\begin{displaymath}
\hspace*{1cm} \left( \begin{array}{ccc}
       c_l        & -s_l & 0\\
       s_l        & c_l & 0\\
       0          & 0   & 1 \end{array} \right) \nonumber \\
\end{displaymath}    
\begin{displaymath}
      = \left( \begin{array}{cc}
          s_{\nu} s_l c + c_{\nu} c_l e^{- i \psi} &
          s_{\nu} c_l c - c_{\nu } s_l e^{-i \psi} \\
          c_{\nu} s_l c - s_{\nu} c_l e^{-i \psi} &
          c_{\nu} c_l c + s_{\nu} s_l e^{-i \psi} \\
          -s_ls & c_ls \end{array} \right. \,  
\end{displaymath}
\begin{equation}
\hspace*{5cm} \left. \begin{array}{c}
         s_{\nu } s\\
         c_{\nu } s\\
         c \end{array} \right) \, .
\end{equation}
The leptonic mixing angles are given by $\theta _l$, describing a mixing for
the charged leptons, an angle $\theta $, describing a mixing between the
second and the third family, and an angle $\theta _{\nu}$, describing the
mixing in the neutrino channel. The complex phase, causing $CP$ violation for
the leptons, is denoted by $\phi $. For simplicity we assume $CP$ symmetry
to be conserved in the leptonic sector.
The electroweak doublets written above give the following leptonic flavor
mixing matrix:
\begin{eqnarray}
V_l & = & \left( \begin{array}{ccc}
\frac{1}{\sqrt{2}}   &   \frac{1}{\sqrt{6}} & \frac{1}{\sqrt{3}} \\
- \frac{1}{\sqrt{2}} &   \frac{1}{\sqrt{6}} & \frac{1}{\sqrt{3}} \\
          0          & - \frac{2}{\sqrt{6}} & \frac{1}{\sqrt{3}} 
          \end{array} \right) \nonumber \\
     & = & \left( \begin{array}{ccc}
        \frac{1}{\sqrt{2}} &   \frac{1}{\sqrt{2}} & 0 \\
      - \frac{1}{\sqrt{2}} &   \frac{1}{\sqrt{2}} & 0 \\
          0                &          0           & 1 
          \end{array} \right) 
     \left( \begin{array}{ccc}
        1   &  0  & 0 \\
        0   &   \frac{1}{\sqrt{3}} & \frac{2}{\sqrt{6}} \\
        0   &  -\frac{2}{\sqrt{6}} & \frac{1}{\sqrt{3}}  
                    \end{array} \right) \nonumber \\
\end{eqnarray}
We can read off the following mixing angles:
\begin{displaymath}
\theta_l = 0, \hspace*{0.5cm} \theta_{\nu} = {\rm arcsin}
\frac{1}{\sqrt{2}} = 45^{\circ},
\end{displaymath}
\begin{equation}
\theta = {\rm arcsin} \frac{2}{\sqrt{6}} = 54.7^{\circ} \, .
\end{equation}
We note that ${\rm sin}^2 \, 2 \theta_{\nu} = 1$ and ${\rm sin}^2 \, 2 \theta
= 8/9$. Using the arguments given in \cite{frixi1} we can also write down
the corrections to the above (lowest--order) leptonic mixing matrix, given
by the violation of the democratic symmetry for the charged leptons. In an
illustrative example, we obtain
\begin{displaymath}
V'_l = V_l + \frac{m_{\mu}}{m_{\tau }}
\left( \begin{array}{ccc}
0 & \frac{1}{\sqrt{6}} & - \frac{1}{2\sqrt{3}} \\
0 & \frac{1}{\sqrt{6}} & - \frac{1}{2\sqrt{3}} \\
0 & \frac{1}{\sqrt{6}} & - \frac{1}{\sqrt{3}}
\end{array} \right)
\end{displaymath}
\begin{equation}
\hspace*{1cm} - \sqrt{\frac{m_e}{m_{\mu}}}
\left( \begin{array}{ccc}
\frac{1}{\sqrt{6}} & - \frac{1}{\sqrt{2}} & 0\\
\frac{1}{\sqrt{6}} &   \frac{1}{\sqrt{2}} & 0\\
- \frac{2}{\sqrt{6}} &         0        & 0 \end{array} \right) \, .
\end{equation}
The corrections to  $\theta_l$ and $\theta $ are small, i. e.
$\theta _l = - 4.1^{\circ}$ and $\theta = 52.3^{\circ}$. The value of
$\theta _{\nu}$ is essentially unchanged. Correspondingly we find
${\rm sin}^2 \, 2 \theta = 0.94$. Similar results for ${\rm sin}^2 2 \theta$
have also been obtained in \cite{fuku}.

We have obtained large mixing angles for all neutrinos. Each flavor
eigenstate $(\nu_e, \nu_{\mu}$ and $\nu_{\tau})$, is a linear superposition
of three mass eigenstates, $\nu_1, \nu_2$ and $\nu_3$,
described by $V'^{\dagger}$. The
electron neutrino is in lowest order given by
\begin{equation}
\nu _e = \frac{1}{\sqrt{2}} \left( \nu_1 - \nu_2 \right) \, .
\end{equation}
An electron neutrino produced in the sun would oscillate between the states
$\nu _e = \left( \nu_1 - \nu_2 \right) / \sqrt{2}$ and
$\stackrel{\sim}{\nu} _e = \left( \nu_1 + \nu_2 \right) / \sqrt{2}$. Note
that in our case
this state is neither a $\mu $--neutrino nor a $\tau $--neutrino, but rather
a mixture of the two. The results of the solar neutrino experiments are
consistent with such a large mixing angle. For example, in the case of
long--wavelength vacuum oscillations one has:
\begin{displaymath}
P \left( \nu_e \rightarrow \nu_e \right) = 1 -
\end{displaymath}
\begin{equation}
{\rm sin}^2 \,
2 \theta_{\rm sun} sin^2
\left( {\rm 1.27} \frac{\Delta m^2_{\rm sun} L}{|{\bf P}|} \right)
\end{equation}
with ${\rm sin}^2 2 \theta_{\rm sun} \approx (0.6 \ldots 1.1) \times 10^{-10}
{\rm eV}^2$ (see \cite{kearns}). In our case we have ${\rm sin}^2
2 \theta_{\rm sun} = {\rm sin}^2 2 \theta_{\nu} = 1$. Then $|m^2_2 - m^2_1|
= \Delta m ^2_{\rm sun} \sim 10^{-10}$ eV$^2$. i. e., the two neutrinos
$\nu_1$ and $\nu_2$ must be degenerate to a very high degree of accuracy.

In terms of mass eigenstates the $\mu$-- and $\tau$--neutrinos are given
approximately by:
\begin{eqnarray}
\nu_{\mu} & = & \frac{1}{\sqrt{6}} \left( \nu_1 + \nu_2 - 2\nu_3 \right) 
\nonumber \\
\nu_{\tau} & = & \frac{1}{\sqrt{3}} \left( \nu_1 + \nu_2 + \nu_3 \right) \, .
\end{eqnarray}
A $\mu$--neutrino will in general oscillate into all three neutrinos. However,
due to the high degeneracy between the $\nu_1$ and $\nu_2$--states,
oscillations between $\mu$--neutrinos and electron--neutrinos will appear
only at very large distances. Oscillations between $\mu$--neutrinos and
$\tau $--neutrinos could show up at smaller distances, if the mass difference
between the $\left( \nu_1, \nu_2 \right) $--states and the $\nu_3$--state is
sizable. For the sake of our discussion let us suppose that the first two
neutrino states are completely degenerate, in which case we can perform a
45$^{\circ}$--rotation among the two states without changing the physical
situation.

The atmospheric neutrino experiments, in particular the recent
Superkamiokande measurements \cite{fuku1}, are consistent with a large
mixing angle for the $\nu_{\mu} \leftrightarrow \nu_{\tau}$ oscillations:
\begin{displaymath}
P \left( \nu_{\mu} \rightarrow \nu_{\tau} \right) =
\end{displaymath}
\begin{equation}
{\rm sin}^2 2 \theta_{\rm atm} {\rm sin}^2
\left( {\rm 1.27} \frac{\Delta m^2_{\rm atm} L}{|P|} \right)
\end{equation}
with ${\rm sin}^2 \, 2 \theta_{\rm atm} \approx (0.7 \ldots 1)$ and
$\Delta m^2_{\rm atm} \approx (0.3 \ldots 8) \times 10^{-3}$ eV$^2$. In our
case we obtain
${\rm sin}^2 \, 2 \theta_{\rm atm} = {\rm sin}^2 2 \theta = {\rm 0.83}$ if we
disregard the small symmetry violations given in eq. (14).

Thus the observational hints towards neutrino oscillations both for solar and
atmospheric neutrinos indicate a mass pattern for the three neutrino states
as follows. The first two neutrinos $\nu_1$ and $\nu_2$ are almost
degenerate, while the mass of the third neutrino $\nu_3$ is slightly larger
or smaller. If this picture is correct, it would imply that there is no way
to obtain evidence for neutrino oscillations using neutrinos beams from
accelerators, unless one performs a long distance experiment.

The pattern for the neutrino mass differences discussed above can be realized
in two qualitatively different ways. Either there is a hierarchical pattern
in the neutrino mass pectrum or the three neutrino masses are nearly
degenerate. If the spectrum is hierarchical, the first two mass eigenstates
$\nu_{1,2}$ must be extremely light. For example one could have
$m_1 \sim 0$ and $m_2 \sim 10^{-5}$ eV. The third neutrino $\nu _3$ would
have a mass in the range (0.02 $\ldots$ 0.1) eV. Thus one has
$\Delta m^2 _{\rm sum} \approx m^2_2$ and $\Delta m^2_{\rm atm} \approx
m^2_3$. In this case the neutrino masses would play only a minor role in
cosmology.

In the case of a degenerate neutrino mass spectrum the three mass
eigenstates would have nearly the same mass:
\begin{eqnarray}
m_1 & = & \delta_{\nu} \, , \nonumber \\
m_2 & = & \delta_{\nu} + \kappa_{\nu} \, ,
\hspace{0.5cm} \left(|\kappa_{\nu} / \delta_{\nu}| \ll 1 \right) \, , \nonumber \\
m_2 & = & \delta_{\nu } + \kappa'_{\nu} \, ,
\hspace*{0.5cm} \left( |\kappa'_{\nu} / \delta_{\nu}| \ll 1 \right) \, ,
\end{eqnarray}
with the constraints
$\Delta m^2_{\rm sun} \approx 2 \delta_{\nu} \kappa_{\nu} \sim
10^{-10}$ eV$^2$ and $\Delta m^2_{\rm atm} \approx 2 \delta _{\nu }
\kappa'_{\nu } \sim 10^{-3}$ eV$^2$. If neutrinos are a significant part of
the dark matter
component in the universe, the sum of the three neutrino states is expected
to be in the range between 4 and 8 eV. As an illustrative example we could
take $\delta_{\nu} =$ 2 eV, i. e., $m_1 = $ 2eV, $m_2 = \left( 2 + 2 \times
10^{-11} \right)$ eV, and $m_3 = \left( 2 + 10^{-3} \right)$ eV.

It remains to be seen whether further experimental studies of the neutrino
mixing phenomena support both the large mixing angles and the mass spectrum
discussed above. Let us finally stress the very simple pattern which arises
if the three massive neutrinos are nearly degenerate. In this case we have,
neglecting the small violations of the democratic symmetry:
\begin{eqnarray}
M_l^- & = & \frac{C_l}{3} \left( \begin{array}{ccc}
                                  1 & 1 & 1 \\
                                  1 & 1 & 1 \\
                                  1 & 1 & 1 \end{array} \right) \nonumber \\
M_{\nu } & = & C_{\nu} \left( \begin{array}{ccc}
                                1 & 0 & 0 \\
                                0 & 1 & 0 \\
                                0 & 0 & 1 \end{array} \right)
\end{eqnarray}
From a different point of view nearly degenerate neutrinos have been
discussed in \cite{ioannis}.

The case of nearly degenerate
neutrinos, possibly of masses of the order of 2 eV, is an interesting
possbility, also from a cosmological point of view. Large mixing angles are
readily obtained in such a model. Both the mass spectrum and the flavor
mixing pattern of the leptons might indeed differ substantially from those
observed for the quarks.

\end{document}